\documentclass[twocol]{ametsoc}
\usepackage{subfigure}

\usepackage{ulem}

\newcommand{\pd}[1]{\partial_{#1}}

\DeclareMathOperator{\cov}{cov}

\journal{mwr}

\bibpunct{(}{)}{;}{a}{}{,}

\title{Improving particle filter performance by smoothing observations}

\authors{Gregor Robinson, Ian Grooms\correspondingauthor{Ian Grooms, Department of Applied Mathematics, University of Colorado, Boulder, Colorado, USA}, and William Kleiber}
\affiliation{Department of Applied Mathematics, University of Colorado, Boulder}
\email{ian.grooms@colorado.edu}

\abstract{This article shows that increasing the observation variance at small scales can reduce the ensemble size required to avoid collapse in particle filtering of spatially-extended dynamics and improve the resulting uncertainty quantification at large scales.
Particle filter weights depend on how well ensemble members agree with observations, and collapse occurs when a few ensemble members receive most of the weight.
Collapse causes catastrophic variance underestimation.
Increasing small-scale variance in the observation error model reduces the incidence of collapse by de-emphasizing small-scale differences between the ensemble members and the observations.
Doing so smooths the posterior mean, though it does not smooth the individual ensemble members.
Two options for implementing the proposed observation error model are described.
Taking discretized elliptic differential operators as an observation error covariance matrix provides the desired property of a spectrum that grows in the approach to small scales.
This choice also introduces structure exploitable by scalable computation techniques, including multigrid solvers and multiresolution approximations to the corresponding integral operator.
Alternatively the observations can be smoothed and then assimilated under the assumption of independent errors, which is equivalent to assuming large errors at small scales.
The method is demonstrated on a linear stochastic partial differential equation, where it significantly reduces the occurrence of particle filter collapse while maintaining accuracy.
It also improves continuous ranked probability scores by as much as 25\%, indicating that the weighted ensemble more accurately represents the true distribution.
The method is compatible with other techniques for improving the performance of particle filters.}

\begin{document}

\maketitle

\section{Introduction}
Particle filters are a class of ensemble-based methods for solving sequential Bayesian estimation problems. 
They are uniquely celebrated due to their provable convergence to the correct posterior distribution in the limit of an infinite number of particles, with minimal constraints on prior and likelihood \citep{CD02}. Processes that are nonlinear and non-Gaussian can be filtered in this flexible framework, with rigorous assurances of asymptotically correct uncertainty quantification. These advantages stand in contrast to ensemble Kalman filters that lack convergence guarantees for nonlinear or non-Gaussian problems, and to variational methods that provide a point estimate but do not quantify uncertainty in the common case where the Hessian of the objective is unavailable.

The simplest form of a particle filter is descriptively called sequential importance sampling (SIS). We briefly describe the algorithm here to fix notation and terminology, and recommend \cite{DdFG01} for a gentler introduction.

SIS begins by approximating the prior probability distribution with density $p(\mathbf{x}_{j-1})$ at discrete time $j-1$ as a weighted ensemble of $N_e$ members $\left\{ \left(\mathbf{x}_{j-1}^{(i)},w_{j-1}^{(i)}\right) \right\}$, where the weights $w_{j-1}^{(i)}$ are related to the prior probabilities of the corresponding states $\mathbf{x}_{j-1}^{(i)}$. The superscript $(i)$ indexes the collection of particles, and the sum of the weights is one.
This kind of approximation, an \emph{importance sample}, is an ensemble drawn from one distribution that is easy to sample and then reweighted to represent another distribution of interest.

The distribution of interest is the Bayesian posterior at discrete time $j$, which is proportional to the product of the prior at time $j-1$, $p\left(\mathbf{x}_{j-1}\right)$, the transition kernel $p(\mathbf{x}_j|\mathbf{x}_{j-1})$, and the likelihood $p(\mathbf{y}_j|\mathbf{x}_j)$.
SIS evolves the samples from time $j-1$ to time $j$ according to a \emph{proposal kernel} that takes the generic form $p\left(\mathbf{x}^{(i)}_j|\mathbf{x}_{0:j-1}^{(i)},\mathbf{y}_j\right)$. The weights are updated to reflect the difference between the proposal kernel and the Bayesian posterior at time $j$:
\begin{align}
	w_{j}^i &\propto w_{j-1}^i \frac{p\left(\mathbf{y}_j|\mathbf{x}_j^{(i)}\right) p \left(\mathbf{x}_{j}^{(i)}|\mathbf{x}_{j-1}^{(i)}\right)}{p\left(\mathbf{x}^{(i)}_j|\mathbf{x}_{0:j-1}^{(i)},\mathbf{y}_j\right)}. \label{eq:posterior_weights}
\end{align}
The proposal kernel is often set to equal the transition kernel, which simplifies the ratio in \eqref{eq:posterior_weights} so that the weights are proportional to the likelihood: $w_{j}^i \propto w_{j-1}^i \cdot p\left(\mathbf{y}_j|\mathbf{x}_j^{(i)}\right)$.
The proportionality constant is chosen so that the weights sum to one.
(Some authors, e.g.~\citet{vanLeeuwen10}, integrate out dependence on $x_{j-1}$; we instead follow the convention of \citet{DdFG01}.)

Despite its attractive qualities, particle filtering is unpopular in meteorological applications due to an especially vexing curse of dimensionality. The problem is that the importance sampling weights associated with system replicates (\emph{particles}) have a tendency to develop \emph{degeneracy} as the system dimension grows. That is to say, a single particle near the observation will have essentially all the sampling weight while the rest of the particles, bearing effectively zero weight, are ignored in the computation of ensemble statistics.

One can quantify the degree of degeneracy with an \emph{effective sample size} (ESS), which is a  heuristic measurement of the importance sample quality defined as
\begin{align}
	\text{ESS}_j &= \frac{1}{\sum_{i=1}^{N_e} \left(w_j^{(i)}\right)^2}. \label{eq:ess}
\end{align}
The ESS ranges from one if a single weight is nonzero (which is the worst case), to $N_e$ if all weights are equal. If the effective sample size becomes much smaller than the ensemble size, the filter is said to have \emph{collapsed}. A simple approach to combat collapse is to resample the particles from time to time, eliminating particles with low weight and replicating particles with high weights. There are several common approaches to resampling \citep[e.g.][]{DJ08}, and by construction of this resampling step, all weights become uniform: $w_j^{(i)} \rightarrow 1/N_e$ [see also the more recent resampling alternatives in \citet{Reich13} and \citet{AdWR17}]. The term `particle filter' commonly implies an SIS filter with a resampling step, also known as Sequential Importance Resampling (SIR).

SIR particle filters are guaranteed to converge to the correct Bayesian posterior in the limit of an infinite number of particles, but the rate of convergence can be prohibitively slow for high-dimensional problems.
The number of particles required to avoid collapse is typically exponential in a quantity related to the number of observations, as described by \cite{BBL08} and \cite{SBBA08}.
For example, consider a system with Gaussian prior on $\mathbf{x}_j$ and with likelihood, conditional on $\mathbf{x}_j$,
\begin{align}  \label{eq:y.normal}
  {\bf y}_j | {\bf x}_j &\sim
  \mathcal{N}({\bf H}{\bf x}_j,{\bf R})
\end{align}
where $\mathcal{N}(\mathbf{\mu},\bm{\Sigma})$ denotes a multivariate normal distribution with mean $\mathbf{\mu}$ and covariance $\bm{\Sigma}$, {\bf H} is a linear observation operator, and {\bf R} is the covariance of the additive observation error.
For this example \cite{SBBA08} show that the number of particles $N_e$ required to avoid collapse is on the order of $\exp \{ \tau^2/2\}$ where
\begin{align}
	\tau^2 &= \sum_{k=1}^{N_y} \lambda_k^2 \left( \frac{3}{2} \lambda_k^2+1 \right), \label{eq:snyder_tau}
\end{align}
in which $N_y$ is the dimension of the observations and $\lambda^2_k$ are eigenvalues of
\begin{align}
	\cov \left( \mathbf{R}^{-1/2} \mathbf{H} \mathbf{x}_j \right). \label{eq:snyder_eigvals}
\end{align}
\citet{CM13} also discuss the notion of `effective dimension' and how it relates to particle filter performance.
\citet{APSS17} give precise, non-asymptotic results on the relationship between the accuracy of the particle filter, the number of particles, and the `effective dimension' of the filtering problem in both finite and infinite dimensional dynamical systems.
For simplicity of exposition we rely on the formulas quoted here from \citet{SBBA08} and \citet{SBM15}.

A number of methods developed to minimize degeneracy in high-dimensional problems utilize a proposal kernel that is different from the transition prior, using observations to guide proposals. Of all possible proposals that depend only on the previous system state and the present observations, there exists an optimal proposal that minimizes both the variance of the weights and the number of particles required to avoid degeneracy \citep{DGA00,SBM15}. It is typically impractical to sample from that optimal proposal. The various methods proposed to minimize weight degeneracy in practice include the implicit particle filter \citep{CT09,CMT10,CT12,MTAC12}, and the equivalent weights particle filter \citep{vanLeeuwen10,AvL13,AvL15}. 
\citet{SBM15} have shown that improved proposals can reduce the number of particles required to avoid collapse, but the number is still prohibitive for meteorological applications.
Another approach to improving the performance of particle filters uses `localization.'
Localization reduces the effective number of observations (and therefore the required number of particles) by breaking the assimilation into a sequence of smaller subsets.
Localization can also improve the performance of particle filters \citep{PM15,RvH15,Poterjoy16,MHS17}, but breaks convergence guarantees.
Other methods improve the filter results by making the observation error model state dependent \citep{OMB14,ZhuEtAl16}.

This paper describes a different but compatible approach for improving the dimensional scaling of particle filters by smoothing observations before proceeding as though the observations are uncorrelated; equivalently, we increase the small-scale variance in the error model. The goal of doing so is to achieve more desirable dimensional scaling. Whereas changing the proposal kernel allows particle filtering to sample a given posterior more efficiently, manipulating the observation model changes the posterior itself. This may seem to vitiate convergence guarantees at least as badly as localization does. After all, it is possible that localized particle filters and EnKFs converge to some distribution in the large ensemble limit. However, convergence results are still an open problem for EnKFs and localized particle filters. In any case, the limiting distribution of a localized filter is not the true Bayesian filter, and the nature of the bias in the limiting distribution is unknown. By contrast, we can guarantee convergence to a surrogate distribution with bias that can be described and controlled.

The key insight motivating our approach is evident in \eqref{eq:snyder_eigvals}: increasing the observation error variance for any eigenvector of {\bf R} correspondingly decreases the number of particles required. The challenge is to make the problem less expensive to sample with a particle filter, while still accurately incorporating observations on the most physically relevant large scales.
This paper describes an analytically transparent and computationally efficient method that reduces the number of particles required to avoid collapse by increasing the observation error variance at small scales.

\section{Theory}
In this section we develop intuition by considering the observation error model (\ref{eq:y.normal}) in the special case where $\mathbf{R}$ and $\cov(\mathbf{x}_j)$ are Fourier diagonalizable and {\bf H} = {\bf I}. Writing eigenvalues of {\bf R} as $\gamma^2_k$ with $k$ an integer wavenumber from $1$ to $N_y$, and the eigenvalues of $\cov(\mathbf{x}_j)$ as $\sigma_k^2$, the matrix in \eqref{eq:snyder_eigvals} has eigenvalues 
\begin{align}
	\lambda^2_k =\sigma^2_k/\gamma^2_k.
\end{align}
The effects of aliasing complicate the Fourier scale analysis of filtering when observations are not available at every grid point, especially when the observation grid is irregular \citep[Chapter 7]{MH12}.

Recall from the introduction that Snyder et al.'s estimate (\ref{eq:snyder_tau}) of the ensemble size required depends on the system covariance, the observing system, and the observation error covariance. Let us ground the theoretical discussion with general comments about the nature of these quantities in operational numerical weather prediction. Typically the model physics are reasonably well-known and held fixed, so we take $\cov(\mathbf{x}_j)$ to be given.\footnote{One can in principle design physical models to make an assimilation problem more tractable to a particle filter, homologous to the approach we describe that alters the observation model. We do not consider that in this article because the theory scantly differs and the praxis is much more problem dependent. The related representation errors, arising from a mismatch between the length scales resolvable by the numerical model and the length scales present in the observations, are difficult to quantify but are presumably spatially correlated.}
The observing system, like the dynamical model, is typically given and fixed.
The observation error covariance, in contrast both to the dynamical model and the observing system, is often a crude heuristic approximation that is easier to modify. Observation error is frequently taken to have no spatial correlation, for example $\mathbf{R} \propto \mathbf{I}$ in the case of distant identical thermometers, in which case $\{\gamma_k\}$ are constant. Otherwise the observation error may have strong spatial correlations, as may be expected of satellite observations biased by a spatially smooth distribution of unobserved atmospheric particulates, in which case $\gamma_k \rightarrow 0$ rapidly for large $k$.

\subsection{Impact of observation error model on number of particles required}
The following hypothetical examples demonstrate how the observation error model can affect the number of particles required for particle filtering. We first use Snyder's asymptotic arguments to estimate the particle filter ensemble size required to reconstruct a Bayesian posterior with a correlated observation error model, whose realizations are continuous with probability one, and contrast this with the ensemble size required under the approximation that observation errors are spatially uncorrelated. Making this approximation decreases the particle filter ensemble size required to reconstruct the Bayesian posterior. This progression is designed set the stage for our method; we show that using a peculiar choice of $\mathbf{R}$, possessing a growing spectrum, naturally extends the approximation of correlated errors with uncorrelated errors. Our method decreases the number of particles required to approximate the posterior regardless of whether the true errors are correlated or uncorrelated.

Fields whose correlations gradually decrease with distance have decaying spectra, i.e.~$\gamma_k^2\to0$ at small scales. This has a detrimental effect on the effective dimensionality of the problem. Suppose, for example, that observation error variances $\gamma^2_k = k^{-4}$ and system covariance $\sigma^2_k = k^{-2}$. Then eigenvalues of \eqref{eq:snyder_eigvals} are $\lambda^2_k = k^{2}$ and 
\begin{align}
	\tau^2 \approx\int_1^{N_y}k^2\left(\frac{3}{2}k^2+1\right)\text{d}k \sim \frac{3}{10} N_y^5
\end{align}
where the sum in (\ref{eq:snyder_tau}) has been approximated by an integral.
In this example the effective dimensionality of the problem increases extremely rapidly as the number of observations grows.
A similar argument can be used to show that if $\sigma_k^2$ decays sufficiently faster than $\gamma_k$ at small scales (large $k$), then the effective dimensionality of the system remains bounded in the continuum limit.

When the spatial correlation of the observation error is unknown, it is not uncommon to use a spatially-uncorrelated (i.e.~diagonal) observation error model. This approximation is also popular because it is computationally convenient in ensemble Kalman filters, where it enables serial assimilation \citep{Houtekamer01,BEM01,Whitaker02}.
For observations with correlated errors, such as swaths of remotely sensed data, approximating the errors as spatially uncorrelated changes the posterior relative to a more accurate observation error model with correlations; the approximation seems to work well enough in practice.
The spatially uncorrelated approximation, compared to error models with continuous realizations, also makes particle filtering easier. When the error is spatially uncorrelated, $\gamma_k^2$ does not decay to zero at small scales.
Repeating the asymptotic argument in the preceding paragraph with constant $\gamma_k^2 = 1$ implies $\lambda_k^2 = k^{-2}$, so 
\begin{align}
	\tau^2 \approx\int_1^{N_y}k^{-2}\left(\frac{3}{2}k^{-2}+1\right)\text{d}k \sim \frac{3}{2}
\end{align}
in the continuum limit.
This illustrates that the number of particles required to avoid collapse can be significantly reduced by changing the spatial correlations in the observation error model, and in practice the filter results are still acceptably accurate.

Our proposal is take this approximation a step further: we let observation error covariance grow without bound in the progression to small scales. This model of the observation error, possessing a spectrum bounded away from zero, is called a \emph{generalized random field} (GRF) and has peculiar properties described in the Appendix. Despite those peculiarities of GRFs which complicate analysis of the continuum limit, the finite dimensional vector of observational errors can be treated as a multivariate Gaussian random vector.

In the following subsections we discuss the impact of this observation error model on the posterior, and various numerical methods for constructing and implementing the associated particle filter. We find the theory to be more intuitive in terms of this covariance framework than working with smoothing operators, but the final subsection will make the equivalence precise.

\subsection{Effect of a generalized random field likelihood on posterior\label{sec:GRF_FX}}
The performance advantage, described above, does not come for free. Changing the observation error model changes the posterior. 
To demonstrate how our choice of error model affects the posterior, consider again a fully Gaussian system for which the system covariance $\cov (\mathbf{x}_j)$ has the same eigenvectors as the presumed observation error covariance $\mathbf{R}$, and where the observation operator is the identity. Let $\sigma_k^2$ be eigenvalues of $\cov (\mathbf{x}_j)$ and $\gamma_k^2$ be eigenvalues of $\mathbf{R}$, indexed by $k$ in the diagonalizing basis with index $k$ increasing towards small scales. 
Let $\hat{\bm{x}}_k$ and $\hat{\bm{y}}_k$ denote the projection of the prior mean and observations onto the $k^\text{th}$ eigenvector, respectively.
Then the posterior mean of $p(\hat{\bm x}_k|\hat{\bm y}_k)$ is 
\begin{align}
	\hat{\bm{x}}_k + \frac{\sigma_k^2}{\sigma_k^2+\gamma_k^2}(\hat{\bm{y}}_k-\hat{\bm{x}}_k).
\end{align}    
In order for the posterior mean to be accurate at large scales, it will be necessary to design an observation error model with realistic variance at large scales; we return to this point in section 2\ref{sec:Construction}.
Clearly, if $\gamma_k^2\to\infty$ at small scales then the posterior mean will equal the prior mean at small scales.
If the filter tends to ignore small-scale information, then the small-scale part of the prior mean will eventually tend towards the climatological small-scale mean, which is often zero since climatological means are often large-scale.
This observation error model can therefore be expected to have a smoothing effect on the posterior mean.

This is the price to be paid for reducing the effective dimensionality of the system, but the price is not too high.
Small scales are inherently less predictable than large scales, so loss of small-scale observational information may not significantly damage the accuracy of forecasts.
Practical implementations will need to balance between ignoring enough observational information to avoid particle collapse and keeping enough to avoid filter divergence (i.e.~the filter wandering away from the true state of the system). 

In the same example as above, the eigenvalues of the posterior covariance are
\begin{align*}
	\xi_k^2 &= \frac{\sigma_k^2\gamma_k^2}{\sigma_k^2+\gamma_k^2}.
\end{align*}
As noted above, in order for the posterior variance to be accurate at large scales, it will be necessary to design an observation error model with realistic variance at large scales.
At small scales we argue that $\xi_k^2$ is small (using the notation $\ll 1$) regardless of the behavior of $\gamma_k^2$.
This is because the state $\bm{x}$ is associated with a viscous fluid model whose solutions should be continuous.
A GRF error model with $1 \ll \gamma_k^2$ will lead to a posterior variance close to the prior variance at small scales: $\xi_k^2 \approx \sigma_k^2\ll1$.
A more realistic error model with $\gamma_k^2\ll1$ will lead to a much smaller posterior variance, but in either case $\xi_k^2\ll1$.
This argument suggests that the GRF approach should not have a detrimental effect on the posterior variance when applied to atmospheric or oceanic dynamics, provided that the observation error variance at large scales is realistic.

\subsection{Constructing GRF Covariances\label{sec:Construction}}
In the context of an SIR particle filter using the standard proposal with a nonlinear observation error model of the form
\begin{align*}
	\bm{y}_j = \bm{H}(\bm{x}_j) + \bm{\eta}_j
\end{align*}
where $\bm{\eta}_j\sim\mathcal{N}(0,\text{\bf R})$ is the observation error, the incremental weights are computed using
\begin{align*}
	w_j^{(i)}\propto w_{j-1}^{(i)}\exp\left\{-\frac{1}{2}\left(\bm{y}_j-\bm{H}(\bm{x}_j^i)\right)^T\text{\bf R}^{-1}\left(\bm{y}_j-\bm{H}(\bm{x}_j^i)\right)\right\}.
\end{align*}
The goal of this section is to describe two methods for defining an observation error covariance {\bf R} that has the increasing variance prescribed above, and that allow for rapid computation of the weights.
First, we will suppose that the true observation error variance is known, and we will scale it out so that we are dealing only with the error correlation matrix.
If ${\bf R}_0$ is a diagonal matrix with elements that are the observational error variances, then we will let
\begin{align*}
	\text{\bf R} = \text{\bf R}_0^{1/2}\text{\bf C R}_0^{1/2}
\end{align*}
and we will model the matrix {\bf C}.

There is a well-known connection between stationary Gaussian random fields and elliptic stochastic partial differential equations \citep{RH05,LRL11} that allows fast approximation of likelihoods. 
Specifically, the inverse of the covariance matrix of a discretized random field can in some cases be identified with the discretization of a self-adjoint elliptic partial differential equation (PDE).
The connection extends in a natural way to generalized Gaussian random fields, with the caveat that the covariance matrix rather than its inverse is identified with the discretization of an elliptic PDE.
For example, the matrix {\bf C} can be constructed as a discretization of the operator
\begin{align}
	\left(1- \ell^2 \Delta \right)^{\kappa}, \label{eq:diff_op}
\end{align}
in which $\Delta$ is the Laplacian operator, $\ell>0$ is a tuning parameter with dimensions of length, and $\kappa>0$ controls the rate of growth of eigenvalues.
Both the continuous differential operator and its discretization have positive spectra with eigenvalues growing in wavenumber.
The parameter $\ell>0$ controls the range of scales with eigenvalues close to 1.
For length scales longer than $\ell$ the eigenvalues are close to 1 and the observation error model is similar to the commonly-used diagonal, uncorrelated observation error model.
The large-scale observation error is correct, meaning that the posterior will also be correct at large scales.
For length scales smaller than $\ell$ the observation error variance grows at a rate determined by $\kappa$, rapidly rolling off the influence of small scales.

Taking the matrix {\bf C} to be a discretization of an elliptic PDE permits efficient application of the inverse, as required in computing the weights, by means of sparse solvers.
It is also possible to construct {\bf C}$^{-1}$ directly as the discretization of the integral operator that corresponds to the inverse of this PDE, also enabling fast algorithms that have no limitation to regular observation grids.
These kinds of methods will be explored more fully elsewhere.

An alternative to the PDE based approach for modeling {\bf C} is to simply smooth the observations.
Let the smoothing operator be a matrix {\bf S}, and the smoothed observations be denoted $\bm{y}_s$.
Then the observation model
\begin{align*}
	\bm{y}_S = \text{\bf SR}_0^{-1/2}\bm{y}_j = \text{\bf SR}_0^{-1/2}\bm{H}(\bm{x}_j) + \bm{\eta}_s
\end{align*}
where the smoothed observation errors are assumed to have independent, unit-variance errors, implies incremental importance weights of the form
\begin{multline*}
w_j^{(i)} \propto w_{j-1}^{(i)}\times\\
\exp\left\{-\frac{1}{2}(\bm{y}_j-\bm{H}(\bm{x}_j^i))^T\text{\bf R}_0^{-1/2}\text{\bf S}^T\text{\bf SR}_0^{-1/2}(\bm{y}_j-\bm{H}(\bm{x}_j^i))\right\}.
\end{multline*}
If a smoothing operator $\mathbf{S}$ is available, this is equivalent to setting {\bf C}$^{-1} =$ {\bf S}$^T${\bf S}.
As long as the smoothing operator leaves large scales nearly unchanged while attenuating small scales, the impact on the effective sample size and on the posterior will be as described in the foregoing subsections.
If it is possible to construct {\bf S} to {\it project} onto a large-scale subspace, it would be equivalent to setting certain eigenvalues of the observation error covariance to infinity.

\section{Experimental Configuration\label{sec:Config}}
To illustrate the effects of a GRF likelihood in a simple example, we apply an SIR particle filter to a 1-dimensional linear stochastic partial differential equation,
\begin{align}
    \frac{du}{dt} &= \left(-b -c \frac{d}{dx} + \nu \frac{d^2}{dx^2} \right) u + F_t,
\end{align}
where $b,c,\nu \in \mathbb{R}^+$ are constant scalars and $F$ is a time-dependent stochastic forcing that is white in time and correlated in space with a form described below.
The domain is periodic, with length 2$\pi$.
Such models have been used to test filtering algorithms by \citet{MH12}.
In Fourier space this model can be represented as the It\^o equation
\begin{align}
    d\hat{u} &= -(b + i k c + \nu k^2)\hat{u} ~ dt + \zeta dW, \label{eq:SPDE}
\end{align}
where $\hat{u}$ is the Fourier coefficient at wavenumber $k$, $\zeta$ is the noise amplitude, and $dW$ is a standard circularly symmetric complex white noise.
The coefficients are $b=1$, $c=2\pi$, and $\nu=1/9$.
To mimic turbulence in many physical models, we choose a stochastic forcing $F_t$ that decays linearly for large wavenumbers. Specifically, let
\begin{align}
	\zeta^2 = 1/(1+|k|) \label{eq:noise_spectrum}
\end{align}
such that the variance of the noise is one half of its maximum at wavenumber 1.
This configuration (11-13) is chosen to possess a fairly limited range of active wavenumbers so that the particle filtering problem is tractable.

The model admits an analytical solution to which we can compare experimental results.
Since the dynamic is linear and Fourier coefficients are independent, it follows that each Fourier mode evolves as an Ornstein-Uhlenbeck process independent of all other modes. This means we can efficiently propagate the system by sampling directly from the Gaussian distribution available in closed form for each Fourier coefficient \citep{Oksendal03}:
\begin{align}
    \hat{u}_{t+\Delta t} &= \hat{u}_t e^{-\theta_k ~ \Delta t} + \zeta \sqrt{\frac{1-e^{-2 \theta_{r,k} \Delta t}}{2\theta_{r,k}}} \chi_t,\label{eq:Doob}
\end{align}
where $\theta_k = d + ikc + \nu k^2$, $\theta_{r,k}$ is the real part of $\theta_k$, and $\chi_t$ is a standard circularly symmetric complex normal random variable.
The initial condition for the experiment is drawn from the stationary distribution, obtained as the limit $\Delta t\to\infty$ in (\ref{eq:Doob}), which for each wavenumber is a circularly symmetric complex normal random number of standard deviation $1 / \sqrt{2(1+|k|)\theta_{r,k}}$.

A particular solution, hereafter called the `true system state' solution is computed at 2048 equally spaced points in the 2$\pi$-periodic spatial domain, and at 101 equally-spaced points in the time interval $[0,4]$ (the initial condition being at $t=0$).
From this solution, synthetic observations are generated at every 32$^\text{nd}$ spatial location (except as otherwise noted) by adding samples from a stationary zero-mean multivariate normal distribution with variance 0.36 and correlations of the form $\exp\{-|\delta/0.06|\}$ where $\delta$ is the distance between observations.
There are thus 64 $\times$ 100 total observations (there are no observations of the initial condition).

The standard deviation of the observational error is 0.6, while the pointwise climatological standard deviation of the system is about 0.8.
This is a very high observational noise level; we set the observational noise this high because the theoretical estimates of the required ensemble size are extremely large for smaller observational noise. Observational noise levels in meteorological applications are not usually this high relative to the climatological variability of the system.
Despite this high level of noise, the observing system is dense enough in space and time that the filter is able to recover an accurate estimate of the system.

The GRF observation error covariance, used only for assimilation, is constructed as the periodic tridiagonal matrix formed by the second-order centered finite difference approximation to the operator $0.36(1-\ell^2\pd{x}^2)$. 
The diagonal elements (the observation error variance) are all $0.36(1 + 2(\ell/\delta)^2)$ where $\delta$ is the distance between observations; the elements corresponding to nearest-neighbor covariances are all $0.36(1 - (\ell/\delta)^2)$.
When $\ell=0$ the observation error covariance is diagonal.
The local observation error variances increase when $\ell$ increases, and the nearest-neighbor covariances decrease and can even become negative.
The eigenvectors of this matrix are discrete Fourier modes.
When $\ell$ increases, the variance increases for all Fourier modes except the constant mode, which remains at this baseline variance $0.36$.
Experiments are run with 101 values of $\ell^2$ equally spaced in the interval $[0,1]$.
The GRF observation error covariance is not used to generate the synthetic observations.

Assimilation experiments are run with an SIR particle filter to test how the GRF observation error model impacts its performance.
An ensemble size of $N_e=400$ is used, except as noted otherwise.
The SIR particle filter is configured to resample using the standard multinomial resampling algorithm \cite{DdFG01}.
The ESS is tracked before resampling.
Resampling reduces the information content of the ensemble by eliminating some particles and replicating others; to avoid unnecessary loss of information, resampling is only performed whenever the effective sample size (ESS) falls below $N_e/2$.

Two quantities are used to evaluate the effect of the GRF error model on the particle filter's performance.
The first is the root mean squared error between the particle filter's posterior mean and the true system state, where the mean is taken over the spatial domain.
The second is the continuous ranked probability score \citep[CRPS]{H00,GR07}.
This measures the accuracy of the posterior distribution associated with the particle filter's weighted ensemble.
The score is non-negative; a score of zero is perfect, and smaller scores are better.
It is more common to compare the RMSE to the ensemble spread, a function of the ensemble covariance trace \citep{fortin14}, but the CRPS is a more precise way to describe the quality of a probabilistic estimate.
The CRPS is computed at every point of the spatial and temporal grid of $2048\times100$ points.
We compute the CRPS for a range of different $N_y \in (16,32,64,128)$ in order to probe the effects of changing the number of observations. All assimilation runs with the same $N_y$ use the same observations.

We will gauge particle filter performance with the GRF likelihood by comparing it to the reference case of a particle filter computed using a spatially-uncorrelated likelihood. In some cases we will also want to compare the particle filter estimate to the true Bayesian posterior. Though one of the main reasons for using a particle filter is that it works in nonlinear, non-Gaussian problems, a benefit of experimenting with a linear Gaussian problem is that the exact solution to the optimal filtering problem can be computed for this comparison using the Kalman filter.
In particular, the Kalman filter provides the exact posterior covariance {\bf P}$_k$,
\begin{align*}
	\text{\bf K}_k &= \text{\bf P}_{k|k-1} \text{\bf H}^{T} \left(\text{\bf R} + \text{\bf H} \text{\bf P}_{k|k-1} \text{\bf H}^{T} \right)^{-1} \\
    \text{\bf P}_k &= (\text{\bf I}-\text{\bf K}_k \text{\bf H}) \text{\bf P}_{k|k-1},
\end{align*}
which allows us to estimate the number of particles required to avoid filter degeneracy a priori (without running the particle filter) using (\ref{eq:snyder_tau}) and (\ref{eq:snyder_eigvals}).
The prior covariance at time $k$ is denoted {\bf P}$_{k|k-1}$ in the above formulas.

\section{Results\label{sec:Results}}
We compute $\tau^2$ from the Kalman filter results at $t=4$, the end of the assimilation window.
This gives an approximation to the steady-state filtering problem because the posterior covariance converges exponentially to a limiting covariance \citep{CC09}.
This process is repeated for each of eleven $\ell^2$ linearly distributed between 0 and 1 and the results are plotted in the first panel of Figure 1. 
Note that the $\ell^2=0$ case is a spatially-uncorrelated observation error model. We observe a dramatic reduction in the theoretical number of particles required to avoid filter collapse.
The theory of \cite{BBL08} and \cite{SBBA08} predicts that the spatially-uncorrelated noise model requires on the order of $10^{26}$ particles to avoid collapse in this simple 1-dimensional PDE with 2048 Fourier modes.
As $\ell^2$ increases from 0 to 1, the number of required particles drops rapidly to about 8,000.
In fact, as shown below, the SIR particle filter performs well with $\ell^2=1$ for an ensemble size of $400$.

Reducing $\tau^2$ by increasing $\ell^2$ is a result of increasing the observation variance, and the chosen form of the surrogate observation error model is designed to increase the variance primarily for small scales while leaving large scales intact. The impact on the posterior is visualized in the second panel of Figure 1. This panel shows the time-average RMSE of the particle filter mean of the first 50 Fourier modes, normalized by the climatological standard deviation of each Fourier coefficient, for $\ell^2 \in (0, 0.04, 0.4)$. Here we observe that increasing $\ell^2$ primarily increases the posterior variance at small scales, as designed.

The distribution of ESS throughout the 100 assimilation cycles is plotted in Figure \ref{fig:ess_vs_ell2} for various values of $\ell^2$. 
The box plots are constructed from the time series of ESS over all 100 assimilation cycles.
In this proxy for the quality of uncertainty quantification achieved by the particle filter, we observe approximately a tenfold increase in median ESS with $\ell^2=0.3$ and a thirty-fold increase in median ESS with $\ell^2=1$ compared to $\ell^2=0$.
The ESS averages only 10--20\% of $N_e$ when $\ell^2=1$, with occasional collapses.
This is not inconsistent with the theory, which requires $N_e$ of about 8000 to avoid collapse, yet still shows the significant improvements from using a GRF likelihood with relatively small ensembles.
The results below suggest that the particle filter can give an accurate probabilistic estimate of the system state even when the ESS is a small percentage of the ensemble size.

Next we consider how the root mean square error (RMSE) of the particle filter posterior mean from the true system state depends on $\ell$.
Figure \ref{fig:rmse_vs_ell2} shows box plots of the RMSE as a function of $\ell^2$.
The box plots are constructed from the RMSE time series for the final 90 assimilation time steps in each experiment.
The RMSE appears fairly insensitive to $\ell^2$.
The median RMSE for all cases remains below the observation error standard deviation of 0.6.
These results demonstrate that the particle filter remains a fairly accurate point estimator -- both when the filter is collapsed while $\ell$ is small, and when the posterior is substantially over-dispersed due to large $\ell$. The Kalman filter using the true observation model, which is the optimal filter in the best case scenario for this problem, achieves a median RMSE of 0.32.

The use of a GRF likelihood clearly reduces the incidence of collapse in the particle filter, with mild detriment to the RMSE.
The RMSE measures a spatially-integrated squared error, which can mask errors at small scales.
The arguments of section 2\ref{sec:GRF_FX} suggest that the GRF posterior mean will be inaccurate primarily at small scales.
We visualize the severity of this effect in Figure 5,
which compares the true state (red) to the posterior mean (blue) and to ensemble members (gray) for four different values of $\ell^2$: $0$ (diagonal error model), $0.2$, $0.4$, and $0.6$.
The ensemble members are shaded according to their weight: weights near 1 yield black lines while weights near 0 yield faint gray lines.
At $\ell^2=0$ there are few ensemble members visible, reflecting the fact that the ESS is small.
Nevertheless, the posterior mean is reasonably close to the true state.
As $\ell^2$ increases, the number of visible ensemble members increases (reflecting increasing ESS), and the posterior mean becomes smoother.
Although the posterior mean at $\ell^2=0.6$ is smoother than the true system state, the individual ensemble members are not overly smooth; they are instantiations of the dynamical model and are, as such, qualitatively similar to the true state.

The foregoing results have shown that the GRF observation error model improves the ESS without substantially damaging the RMSE, and that the posterior mean is smoother than the true state but the individual ensemble members (particles) are not too smooth.
We finally test whether the uncertainty quantification afforded by the particle filter is improved by using a GRF observation error model.
To this end we compute the CRPS at each point of the spatio-temporal grid of 2048 $\times$ 100 points.
The median CRPS is computed using all 204,800 spatio-temporal grid points for 101 values of $\ell^2$ equally spaced between 0 and 1.
The result is shown in Fig.~\ref{fig:CRPS}.
Median CRPS with $N_y=64$ improves from about 0.27 at $\ell^2=0$ to 0.22 at $\ell^2=0.3$, and then remains steady or slightly increases at larger $\ell^2$.\footnote{For comparison, the ensemble spread simultaneously improves by a factor of about 2, going from a time-averaged 36\% of RMSE when $\ell^2=0$ to 71\% RMSE when $\ell^2=1$.}
Some sampling variability is still evident in the median CRPS, with occasional values as low as 0.21.

Varying the number of observations, also shown in Figure \ref{fig:CRPS}, displays additional interesting behavior about the distributional estimate the particle filter provides. In each $N_{y}$ case we explored, there is a choice of $\ell^2$ that improves the particle filter CRPS. The differences in optimal $\ell^2$ emphasizes that the optimal parameter depends not only on the active scales in the underlying physics, but also on the resolution of the data.

There is less information to spare about physically important scales when observations are sparse (cf. $N_{y}=16$), in which case there is only a narrow window of suitable choices for $\ell^2 \approx 0.12$ before the smoothing effect deteriorates the predictive quality of the particle filter by over-suppressing active scales in the observations.

On the other hand, dense observations provide more abundant small-scale information that makes the particle filtration more susceptible to collapse. This necessitates a larger choice of $\ell^2$ to achieve optimal particle filter performance. In this case, the more abundant information in denser observations can compensate for the injury we do to the surrogate posterior by more aggressively smoothing away small scales. Indeed the most dramatic improvement in the particle filter's uncertainty quantification occurs for $N_y = 128$. Here the particle filter greatly struggles for small $\ell^2$, where we observe a CRPS over 0.29; however when $\ell^2 \approx 0.7$ the CRPS dips under 0.22, competitive with that of all other observation models considered here. This suggests that smoothing is particularly helpful in improving the particle filter's overall probabilistic estimate when observations are dense.

The CRPS results show that the particle filter's uncertainty quantification is improved by the GRF likelihood: a 25\% decrease (improvement) in CRPS is comparable to the improvement achieved by various statistical post-processing techniques for ensemble forecasts \citep{KRBGMG11,KRG11,SB14,FST15}.
Somewhat surprisingly, the CRPS significantly improves moving from $\ell^2=0$ to $\ell^2=0.1$ despite the fact that the ESS remains quite small.
Overall, these CRPS results suggest that even small improvements in ESS can substantially improve the quality of the probabilistic state estimate.
They also confirm that improving the ESS due to increasing $\ell^2$ must be considered in balance against the consequent departure from the true posterior; the CRPS does not improve at large $\ell^2$, even though the ESS improves, because the surrogate posterior becomes less realistic.

Figure 6 demonstrates how SIR uncertainty quantification depends on ensemble size. The figure shows a kernel density estimate of CRPS over all 2048 grid points and all 100 timesteps, for varying number of particles $N_p \in (100, 200, 400, 800, 1600)$. The CRPS mode remains unchanged, but the mean decreases as the  distribution concentrates around the mode primarily at the expense of mass in the tail. The weak dependence of CRPS on ensemble size underscores the appeal of improving UQ by other means.

\section{Conclusions\label{sec:Conclusions}}
We have demonstrated theoretically (in the framework of \citet{BBL08} and \cite{SBBA08}) and in a simple experiment that the number of particles required to avoid collapse in a particle filter can be significantly reduced through a judicious construction of the observation error model.
This observation error model has large observation error variance at small scales, which reduces the effective dimensionality and focuses attention on the more dynamically-relevant large scales.
This observation error model is equivalent to smoothing observations before proceeding as though the observations are uncorrelated.
The cost of this approach is that it alters the posterior, leading to a smoother posterior mean.
In practice, a balance will need to be found between avoiding collapse and retaining as much observational information as possible.

An observation error model whose variance increases at small scales is associated with a so-called generalized random field (GRF).
This connection allows for rapidly applying the covariance matrix's inverse (which is required to compute the particle weights) using fast numerical methods for self-adjoint elliptic partial differential equations.
The method can also be implemented by smoothing the observations before assimilating them, and then assimilating the smoothed observations with an assumption of independent errors.
Both of these avenues are amenable to serial processing of observations, as required by certain parallel implementations \citep[e.g.][]{AC07}.
All of these approaches are compatible with periodic or aperiodic domains.

The results of the one-dimensional stochastic partial differential equation show that this approach improves the `effective sample size' (ESS), which measures how well the weights are balanced between the particles, by an order of magnitude.
The root mean squared error of the particle filter's posterior mean is not significantly impacted by the approach.
One of the main motivations for using particle filters is that they provide meaningful uncertainty estimates even in problems with nonlinear dynamics and observations, and non-Gaussian distributions.
Thus, the continuous ranked probability score (CRPS) is used to test the quality of the particle filter's associated probability distribution.
The GRF observation error model improves the CRPS by as much as 25\%, which is a large improvement, comparable to results obtained by statistical post-processing of the ensemble \citep[e.g.][]{KRBGMG11,KRG11,SB14,FST15}.
This improvement in CRPS is obtained even when the effective sample size (ESS) is less than 20 out of 400, which shows that good probabilistic state estimation can be achieved even with ESS much less than the ensemble size.
The theoretical results suggest that an ensemble size on the order of 8000 is required to avoid collapse in this example problem.
Good results are obtained with an ensemble size of 400, even though the ensemble does collapse from time to time.

The theory of \citet{SBBA08} estimates the ensemble size required to avoid collapse, which is unrealistically large for typical meteorological applications using standard observation error models.
Using a GRF observation error model increases the ESS for a fixed ensemble size, making it easier to achieve the goal of avoiding collapse.
The approach advocated here may still prove insufficient to enable particle filtering of weather, ocean, and climate problems; the minimum required ensemble size will be reduced, but may still be impractically large.
Happily, the method is entirely compatible with approaches based on altered proposals \citep{CT09,vanLeeuwen10,AvL15} and with localization methods \citep{PM15,RvH15,Poterjoy16}.
The method is also compatible with ensemble Kalman filters and with variational methods, but it is not clear whether the approach would yield any benefit there.

Indeed, dynamics of extratropical synoptic scales are often assumed to be approximately linear and are easily estimated with an ensemble Kalman filter. 
But ensemble Kalman filters do not provide robust uncertainty quantification in the face of nonlinear observation operators or nonlinear dynamics, e.g.~at synoptic scales in the tropics.
In contrast, the method proposed here has the potential to provide robust uncertainty quantification even with nonlinear dynamics and observations.
However, it is still unknown in what contexts our peculiar error model damages the posterior more severely than approximating the system as linear and Gaussian for the sake of assimilating data with ensemble Kalman filters.
We expect performance comparison to be context-dependent, and hope future work will help reveal how to balance advantages and disadvantages that are relevant in practice.

\acknowledgments
The authors are grateful for discussions with C.~Snyder and J.~L.~Anderson, both of whom suggested a connection to smoothing observations, and to the reviewers who suggested numerous improvements.
G.~Robinson was supported by an Innovative Seed Grant from the University of Colorado.
This work used the Extreme Science and Engineering Discovery Environment (XSEDE), which is supported by National Science Foundation grant number ACI-1548562 \citep{XSEDE}. Specifically, it used the Bridges system, which is supported by NSF award number ACI-1445606, at the Pittsburgh Supercomputing Center (PSC) through allocation ATM160010 \citep{Bridges}.

\vfill\null

%
\appendix[]
\appendixtitle{Generalized Random Fields}
Generalized random fields (GRFs) are discussed at length in \cite{Yaglom87}, and a few extra details can be found in \cite{GV64}.
A GRF whose Fourier spectrum is not integrable at small scales has infinite variance.
The prototypical example is a spatially-uncorrelated field, whose spectrum is flat.

A GRF is not defined pointwise.
Rather than being defined pointwise, or `indexed by spatial location,' it is indexed by rapidly decaying test functions (often taken to be elements of a Schwartz space).
This is perhaps best explained by reference to an ordinary random field.
If $Z(\bm{x})$ is a random field that is defined pointwise and $\phi(\bm{x})$ is a test function then we can define a new, `function indexed' random field $Z(\phi)$ using the expression
\begin{align*}
	Z(\phi) = \int Z(\bm{x})\phi(\bm{x})\text{d}x.
\end{align*}
If the field $Z$ is not defined pointwise, it may still be indexed by test functions.

The concept of a covariance function for an ordinary random field can be generalized to a GRF.
The resulting object is a `covariance kernel' which can be a generalized function, i.e.~an element of the dual of a Schwartz space.
The prototypical covariance kernel is the so-called Dirac delta function which is not, in fact, a function.

The observation error covariance model advocated in this article can be conceptualized in two ways.
It can be thought of as an approximation to a GRF where the spectrum has been truncated at the smallest resolvable scale on the grid.
Alternatively, one can assume that observations are not taken at infinitesimal points in space, but rather that the observing instrument senses over a small region of space via some test function $\phi$.
The value of the GRF for an observation is thus indexed by the allowed test functions $\phi$ rather than the spatial location of the observation.

\vfill\null


%
%




%
%
%
\bibliographystyle{ametsoc2014}
\bibliography{references}

\begin{thebibliography}{44}
\providecommand{\natexlab}[1]{#1}
\providecommand{\url}[1]{\texttt{#1}}
\renewcommand{\UrlFont}{\rmfamily}
\providecommand{\urlprefix}{URL }
\expandafter\ifx\csname urlstyle\endcsname\relax
  \providecommand{\doi}[1]{doi:\discretionary{}{}{}#1}\else
  \providecommand{\doi}{doi:\discretionary{}{}{}\begingroup
  \urlstyle{rm}\Url}\fi
\providecommand{\eprint}[2][]{\url{#2}}

\bibitem[{Acevedo et~al.(2017)Acevedo, de~Wiljes,, and Reich}]{AdWR17}
Acevedo, W., J.~de~Wiljes, and S.~Reich, 2017: Second-order accurate ensemble
  transform particle filters. \textit{SIAM J Sci Comput}, \textbf{39~(5)},
  A1834--A1850.

\bibitem[{Ades and Van~Leeuwen(2013)Ades, and Van~Leeuwen}]{AvL13}
Ades, M., and P.~J. Van~Leeuwen, 2013: An exploration of the equivalent weights
  particle filter. \textit{Quart.\ J.\ Roy.\ Meteor.\ Soc.},
  \textbf{139~(672)}, 820--840.

\bibitem[{Ades and Van~Leeuwen(2015)Ades, and Van~Leeuwen}]{AvL15}
Ades, M., and P.~J. Van~Leeuwen, 2015: The equivalent-weights particle filter
  in a high-dimensional system. \textit{Quart.\ J.\ Roy.\ Meteor.\ Soc.},
  \textbf{141~(687)}, 484--503.

\bibitem[{Agapiou et~al.(2017)Agapiou, Papaspiliopoulos, Sanz-Alonso,, and
  Stuart}]{APSS17}
Agapiou, S., O.~Papaspiliopoulos, D.~Sanz-Alonso, and A.~Stuart, 2017:
  Importance sampling: Intrinsic dimension and computational cost.
  \textit{Statistical Science}, \textbf{32~(3)}, 405--431.

\bibitem[{Anderson and Collins(2007)Anderson, and Collins}]{AC07}
Anderson, J.~L., and N.~Collins, 2007: Scalable implementations of ensemble
  filter algorithms for data assimilation. \textit{J Atmos Ocean Tech},
  \textbf{24~(8)}, 1452--1463.

\bibitem[{Bengtsson et~al.(2008)Bengtsson, Bickel,, and Li}]{BBL08}
Bengtsson, T., P.~Bickel, and B.~Li, 2008: \textit{Curse-of-dimensionality
  revisited: Collapse of the particle filter in very large scale systems},
  Collections, Vol. Volume 2, 316--334. Institute of Mathematical Statistics,
  Beachwood, Ohio, USA, \doi{10.1214/193940307000000518},
  \urlprefix\url{http://dx.doi.org/10.1214/193940307000000518}.

\bibitem[{Bishop et~al.(2001)Bishop, Etherton,, and Majumdar}]{BEM01}
Bishop, C.~H., B.~J. Etherton, and S.~J. Majumdar, 2001: Adaptive sampling with
  the ensemble transform kalman filter. part i: Theoretical aspects.
  \textit{Monthly weather review}, \textbf{129~(3)}, 420--436.

\bibitem[{Chorin et~al.(2010)Chorin, Morzfeld,, and Tu}]{CMT10}
Chorin, A., M.~Morzfeld, and X.~Tu, 2010: Implicit particle filters for data
  assimilation. \textit{Comm App Math Com Sc}, \textbf{5~(2)}, 221--240.

\bibitem[{Chorin and Morzfeld(2013)Chorin, and Morzfeld}]{CM13}
Chorin, A.~J., and M.~Morzfeld, 2013: Conditions for successful data
  assimilation. \textit{J.\ Geophys.\ Res.}, \textbf{118~(20)}.

\bibitem[{Chorin and Tu(2009)Chorin, and Tu}]{CT09}
Chorin, A.~J., and X.~Tu, 2009: Implicit sampling for particle filters.
  \textit{Proc.\ Natl.\ Acad.\ Sci.\ (USA)}, \textbf{106~(41)},
  17\,249--17\,254.

\bibitem[{Chorin and Tu(2012)Chorin, and Tu}]{CT12}
Chorin, A.~J., and X.~Tu, 2012: An iterative implementation of the implicit
  nonlinear filter. \textit{ESAIM-Math Model Num}, \textbf{46~(3)}, 535--543.

\bibitem[{Chui and Chen(2009)Chui, and Chen}]{CC09}
Chui, C., and G.~Chen, 2009: \textit{Kalman Filtering}. 4th ed., Springer.

\bibitem[{Crisan and Doucet(2002)Crisan, and Doucet}]{CD02}
Crisan, D., and A.~Doucet, 2002: A survey of convergence results on particle
  filtering methods for practitioners. \textit{IEEE T signal proces},
  \textbf{50~(3)}, 736--746.

\bibitem[{Doucet et~al.(2001)Doucet, De~Freitas,, and Gordon}]{DdFG01}
Doucet, A., N.~De~Freitas, and N.~Gordon, 2001: An introduction to sequential
  {M}onte {C}arlo methods. \textit{Sequential Monte Carlo methods in practice},
  Springer, 3--14.

\bibitem[{Doucet et~al.(2000)Doucet, Godsill,, and Andrieu}]{DGA00}
Doucet, A., S.~Godsill, and C.~Andrieu, 2000: On sequential {M}onte {C}arlo
  sampling methods for {B}ayesian filtering. \textit{Stat\ Comput},
  \textbf{10~(3)}, 197--208.

\bibitem[{Doucet and Johansen(2009)Doucet, and Johansen}]{DJ08}
Doucet, A., and A.~M. Johansen, 2009: A tutorial on particle filtering and
  smoothing: Fifteen years later. \textit{in Oxford Handbook of Nonlinear
  Filtering}, University Press.

\bibitem[{Feldmann et~al.(2015)Feldmann, Scheuerer,, and
  Thorarinsdottir}]{FST15}
Feldmann, K., M.~Scheuerer, and T.~L. Thorarinsdottir, 2015: {Spatial
  postprocessing of ensemble forecasts for temperature using nonhomogeneous
  Gaussian regression}. \textit{Mon.\ Wea.\ Rev.}, \textbf{143~(3)}, 955--971.

\bibitem[{Fortin et~al.(2014)Fortin, Abaza, Anctil,, and Turcotte}]{fortin14}
Fortin, V., M.~Abaza, F.~Anctil, and R.~Turcotte, 2014: Why should ensemble
  spread match the rmse of the ensemble mean? \textit{Journal of
  Hydrometeorology}, \textbf{15~(4)}, 1708--1713.

\bibitem[{Gelfand and Vilenkin(1964)Gelfand, and Vilenkin}]{GV64}
Gelfand, I., and N.~Vilenkin, 1964: \textit{Generalized functions, volume 4:
  Applications of Harmonic Analysis}. AMS Chelsea Publishing.

\bibitem[{Gneiting and Raftery(2007)Gneiting, and Raftery}]{GR07}
Gneiting, T., and A.~E. Raftery, 2007: {Strictly proper scoring rules,
  prediction, and estimation}. \textit{J\ Am\ Stat\ Assoc}, \textbf{102},
  359--378.

\bibitem[{Hersbach(2000)}]{H00}
Hersbach, H., 2000: Decomposition of the continuous ranked probability score
  for ensemble prediction systems. \textit{Weather Forecast}, \textbf{15},
  559--570.

\bibitem[{Houtekamer and Mitchell(2001)Houtekamer, and Mitchell}]{Houtekamer01}
Houtekamer, P.~L., and H.~L. Mitchell, 2001: A sequential ensemble kalman
  filter for atmospheric data assimilation. \textit{Monthly Weather Review},
  \textbf{129~(1)}, 123--137.

\bibitem[{Kleiber et~al.(2011{\natexlab{a}})Kleiber, Raftery, Baars, Gneiting,
  Mass,, and Grimit}]{KRBGMG11}
Kleiber, W., A.~E. Raftery, J.~Baars, T.~Gneiting, C.~F. Mass, and E.~Grimit,
  2011{\natexlab{a}}: Locally calibrated probabilistic temperature forecasting
  using geostatistical model averaging and local bayesian model averaging.
  \textit{Mon.\ Wea.\ Rev.}, \textbf{139~(8)}, 2630--2649.

\bibitem[{Kleiber et~al.(2011{\natexlab{b}})Kleiber, Raftery,, and
  Gneiting}]{KRG11}
Kleiber, W., A.~E. Raftery, and T.~Gneiting, 2011{\natexlab{b}}: Geostatistical
  model averaging for locally calibrated probabilistic quantitative
  precipitation forecasting. \textit{J Am Stat Assoc}, \textbf{106~(496)},
  1291--1303.

\bibitem[{Lindgren et~al.(2011)Lindgren, Rue,, and Lindstr{\"o}m}]{LRL11}
Lindgren, F., H.~Rue, and J.~Lindstr{\"o}m, 2011: {An explicit link between
  {G}aussian fields and {G}aussian {M}arkov random fields: the stochastic
  partial differential equation approach}. \textit{J Roy Stat Soc B},
  \textbf{73~(4)}, 423--498.

\bibitem[{Majda and Harlim(2012)Majda, and Harlim}]{MH12}
Majda, A.~J., and J.~Harlim, 2012: \textit{Filtering complex turbulent
  systems}. Cambridge University Press.

\bibitem[{Morzfeld et~al.(2017)Morzfeld, Hodyss,, and Snyder}]{MHS17}
Morzfeld, M., D.~Hodyss, and C.~Snyder, 2017: What the collapse of the ensemble
  kalman filter tells us about particle filters. \textit{Tellus A},
  \textbf{69~(1)}, 1283\,809.

\bibitem[{Morzfeld et~al.(2012)Morzfeld, Tu, Atkins,, and Chorin}]{MTAC12}
Morzfeld, M., X.~Tu, E.~Atkins, and A.~J. Chorin, 2012: A random map
  implementation of implicit filters. \textit{J Comput Phys}, \textbf{231~(4)},
  2049--2066.

\bibitem[{Nystrom et~al.(2015)Nystrom, Levine, Roskies,, and Scott}]{Bridges}
Nystrom, N.~A., M.~J. Levine, R.~Z. Roskies, and J.~R. Scott, 2015: Bridges: A
  uniquely flexible hpc resource for new communities and data analytics.
  \textit{Proceedings of the 2015 XSEDE Conference: Scientific Advancements
  Enabled by Enhanced Cyberinfrastructure}, ACM, New York, NY, USA, 30:1--30:8,
  XSEDE '15, \doi{10.1145/2792745.2792775},
  \urlprefix\url{http://doi.acm.org/10.1145/2792745.2792775}.

\bibitem[{Okamoto et~al.(2014)Okamoto, McNally,, and Bell}]{OMB14}
Okamoto, K., A.~McNally, and W.~Bell, 2014: Progress towards the assimilation
  of all-sky infrared radiances: an evaluation of cloud effects.
  \textit{Quart.\ J.\ Roy.\ Meteor.\ Soc.}, \textbf{140~(682)}, 1603--1614.

\bibitem[{{\O}ksendal(2003)}]{Oksendal03}
{\O}ksendal, B., 2003: \textit{Stochastic differential equations}. 6th ed.,
  Springer.

\bibitem[{Penny and Miyoshi(2016)Penny, and Miyoshi}]{PM15}
Penny, S.~G., and T.~Miyoshi, 2016: A local particle filter for
  high-dimensional geophysical systems. \textit{Nonlinear Proc Geoph},
  \textbf{23~(6)}, 391--405.

\bibitem[{Poterjoy(2016)}]{Poterjoy16}
Poterjoy, J., 2016: A localized particle filter for high-dimensional nonlinear
  systems. \textit{Mon.\ Wea.\ Rev.}, \textbf{144~(1)}, 59--76.

\bibitem[{Rebeschini and Van~Handel(2015)Rebeschini, and Van~Handel}]{RvH15}
Rebeschini, P., and R.~Van~Handel, 2015: Can local particle filters beat the
  curse of dimensionality? \textit{Ann Appl Probab}, \textbf{25~(5)},
  2809--2866.

\bibitem[{Reich(2013)}]{Reich13}
Reich, S., 2013: A nonparametric ensemble transform method for {B}ayesian
  inference. \textit{SIAM J Sci Comput}, \textbf{35~(4)}, A2013--A2024.

\bibitem[{Rue and Held(2005)Rue, and Held}]{RH05}
Rue, H., and L.~Held, 2005: \textit{{{G}aussian {M}arkov random fields: theory
  and applications}}. CRC press.

\bibitem[{Scheuerer and B{\"u}ermann(2014)Scheuerer, and B{\"u}ermann}]{SB14}
Scheuerer, M., and L.~B{\"u}ermann, 2014: Spatially adaptive post-processing of
  ensemble forecasts for temperature. \textit{J Roy Stat Soc C},
  \textbf{63~(3)}, 405--422.

\bibitem[{Snyder et~al.(2008)Snyder, Bengtsson, Bickel,, and Anderson}]{SBBA08}
Snyder, C., T.~Bengtsson, P.~Bickel, and J.~Anderson, 2008: Obstacles to
  high-dimensional particle filtering. \textit{Mon.\ Wea.\ Rev.},
  \textbf{136~(12)}, 4629--4640.

\bibitem[{Snyder et~al.(2015)Snyder, Bengtsson,, and Morzfeld}]{SBM15}
Snyder, C., T.~Bengtsson, and M.~Morzfeld, 2015: Performance bounds for
  particle filters using the optimal proposal. \textit{Mon.\ Wea.\ Rev.},
  \textbf{143~(11)}, 4750--4761.

\bibitem[{Towns et~al.(2014)}]{XSEDE}
Towns, J., and Coauthors, 2014: {XSEDE: Accelerating Scientific Discovery}.
  \textbf{16}, 62--74.

\bibitem[{van Leeuwen(2010)}]{vanLeeuwen10}
van Leeuwen, P.~J., 2010: Nonlinear data assimilation in geosciences: an
  extremely efficient particle filter. \textit{Quart.\ J.\ Roy.\ Meteor.\
  Soc.}, \textbf{136~(653)}, 1991--1999.

\bibitem[{Whitaker and Hamill(2002)Whitaker, and Hamill}]{Whitaker02}
Whitaker, J.~S., and T.~M. Hamill, 2002: Ensemble data assimilation without
  perturbed observations. \textit{Monthly Weather Review}, \textbf{130~(7)},
  1913--1924.

\bibitem[{Yaglom(1987)}]{Yaglom87}
Yaglom, A.~M., 1987: \textit{Correlation theory of stationary and related
  random functions}. Springer.

\bibitem[{Zhu et~al.(2016)}]{ZhuEtAl16}
Zhu, Y., and Coauthors, 2016: All-sky microwave radiance assimilation in
  ncep’s gsi analysis system. \textit{Mon.\ Wea.\ Rev.}, \textbf{144~(12)},
  4709--4735.

\end{thebibliography}

%

%

\begin{figure*}
\begin{subfigure}
	\centering
    \includegraphics[width=19pc]{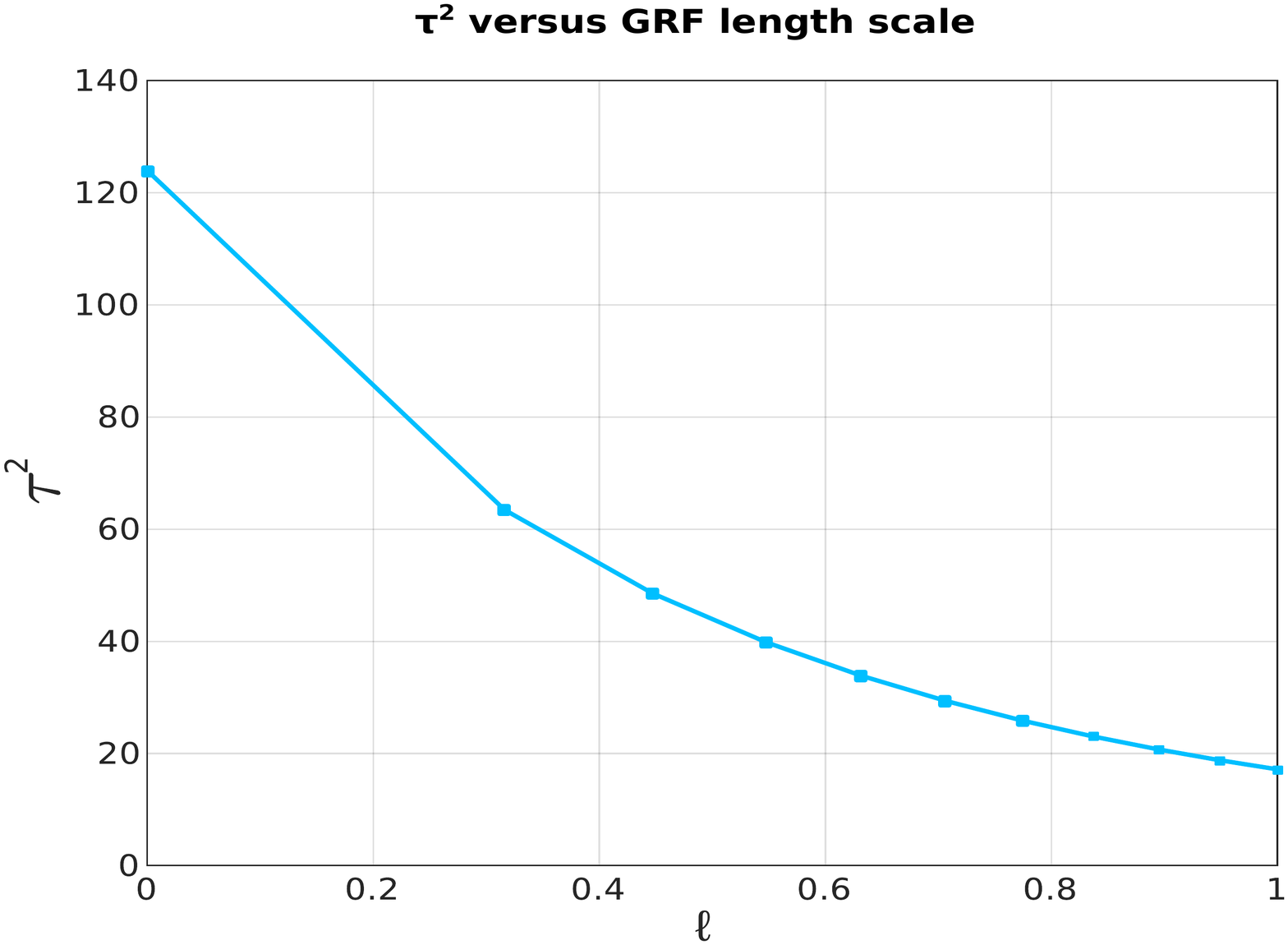}
    \label{fig:tau2_vs_ell}
\end{subfigure}
\begin{subfigure}
	\centering
    \includegraphics[width=19pc]{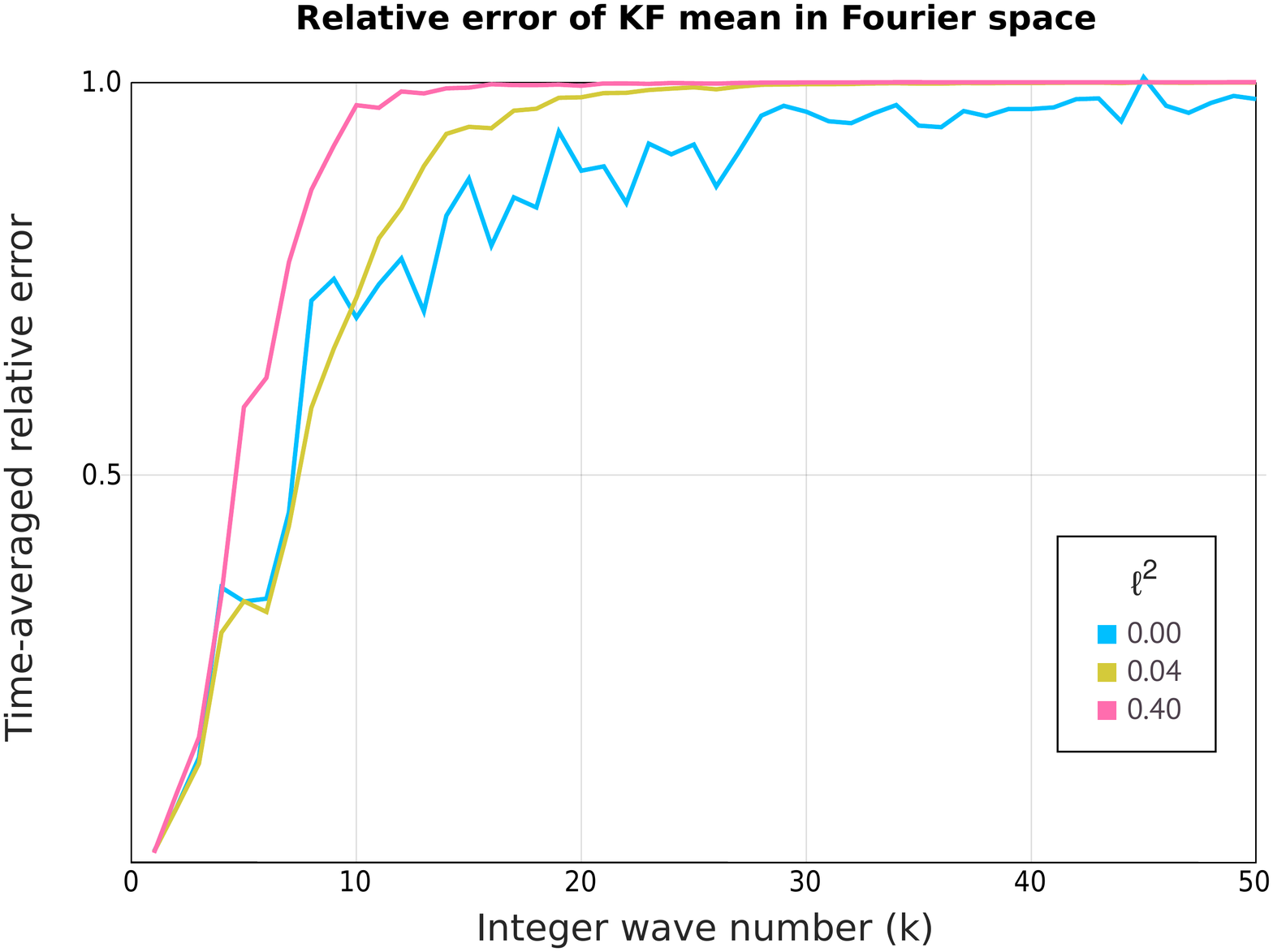}
    \label{fig:relative_error_fourier}
\end{subfigure}
    \caption{The left panel shows $\tau^2$ \eqref{eq:snyder_tau} for different values of GRF length scale $\ell$. Because the number of particles required to avoid degeneracy increases exponentially in $\tau^2/2$, the observed decrease in $\tau^2$ as we roll off scales greater than $\ell$ indicates a reduced computational burden in using particle filtering for uncertainty quantification. Similarly, the decrease suggests that for fixed computation cost one may be able to mitigate the variance underestimation that tends to plague particle filters in high dimensions. Although the ordinate in this figure is $\ell$ to make direct contact with the length scale, all other figures are given in terms of $\ell^2$ to relate more directly to the spectrum of the GRF likelihood.
    The panel on the right shows the RMS error in the Kalman Filter’s posterior mean, in Fourier space, normalized by the climatological standard deviation of each Fourier coefficient for different values of $\ell^2$. Here we see how the error in the posterior mean, considered as a function of wavenumber, approaches the climatological standard deviation more rapidly when $\ell^2$ is larger. It is exactly this posterior variance increase at small scales that underpins our approach: a posterior with larger total variance is easier for a particle filter to sample, while keeping the posterior accurate at large scales is key in forecast.}
\end{figure*}

\begin{figure*}
	\centering
    \includegraphics[width=\textwidth]{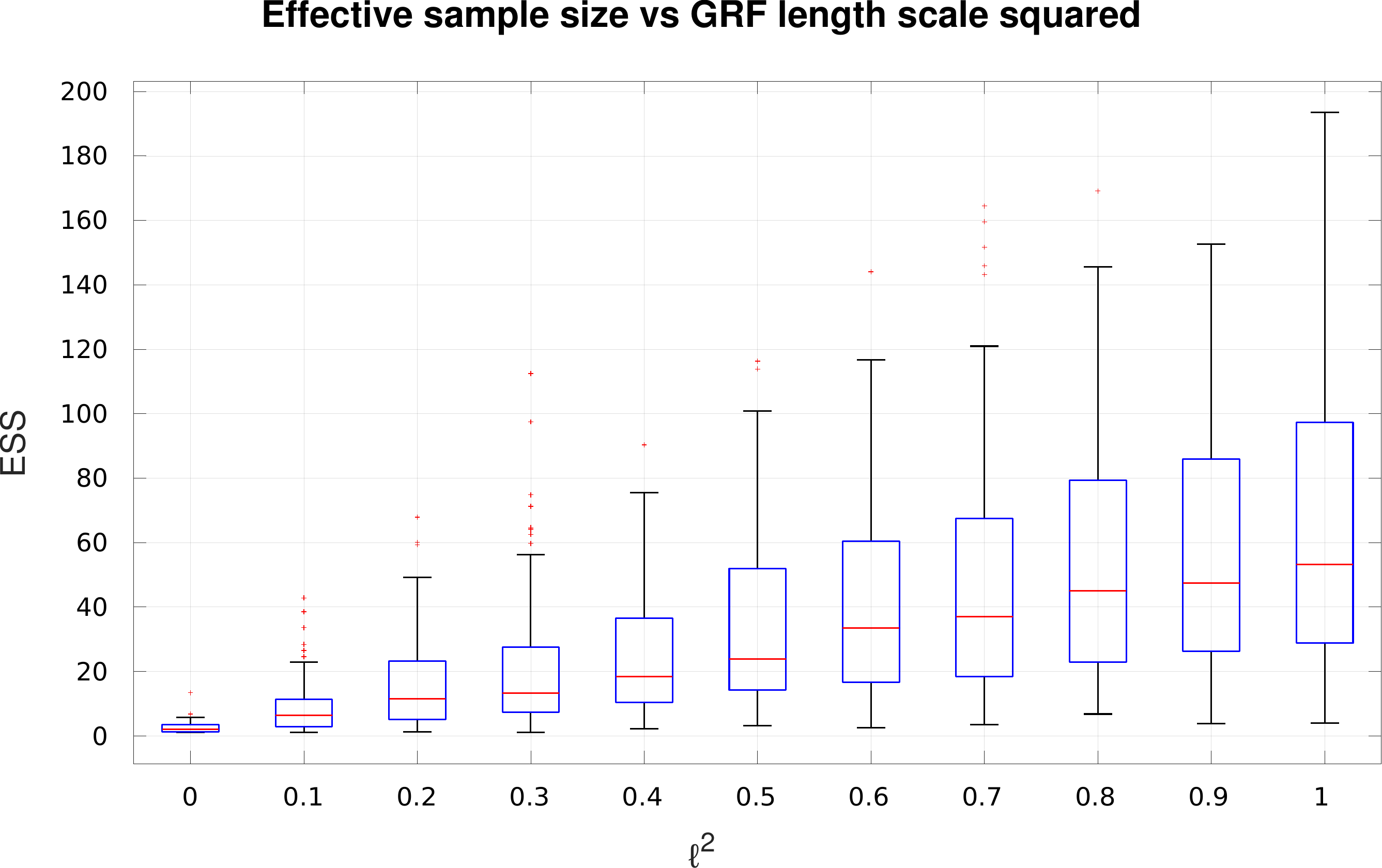}
    \caption{Effective sample size \eqref{eq:ess} distributions for different values of $\ell^2$ from $0$ to $1$. Each box represents the middle 50\% quantile, a central line representing the median, and the whiskers span the data not considered outliers by the 1.5$\times$IQR rule.}\label{fig:ess_vs_ell2}
\end{figure*}

\begin{figure*}
	\centering
    \includegraphics[width=\textwidth]{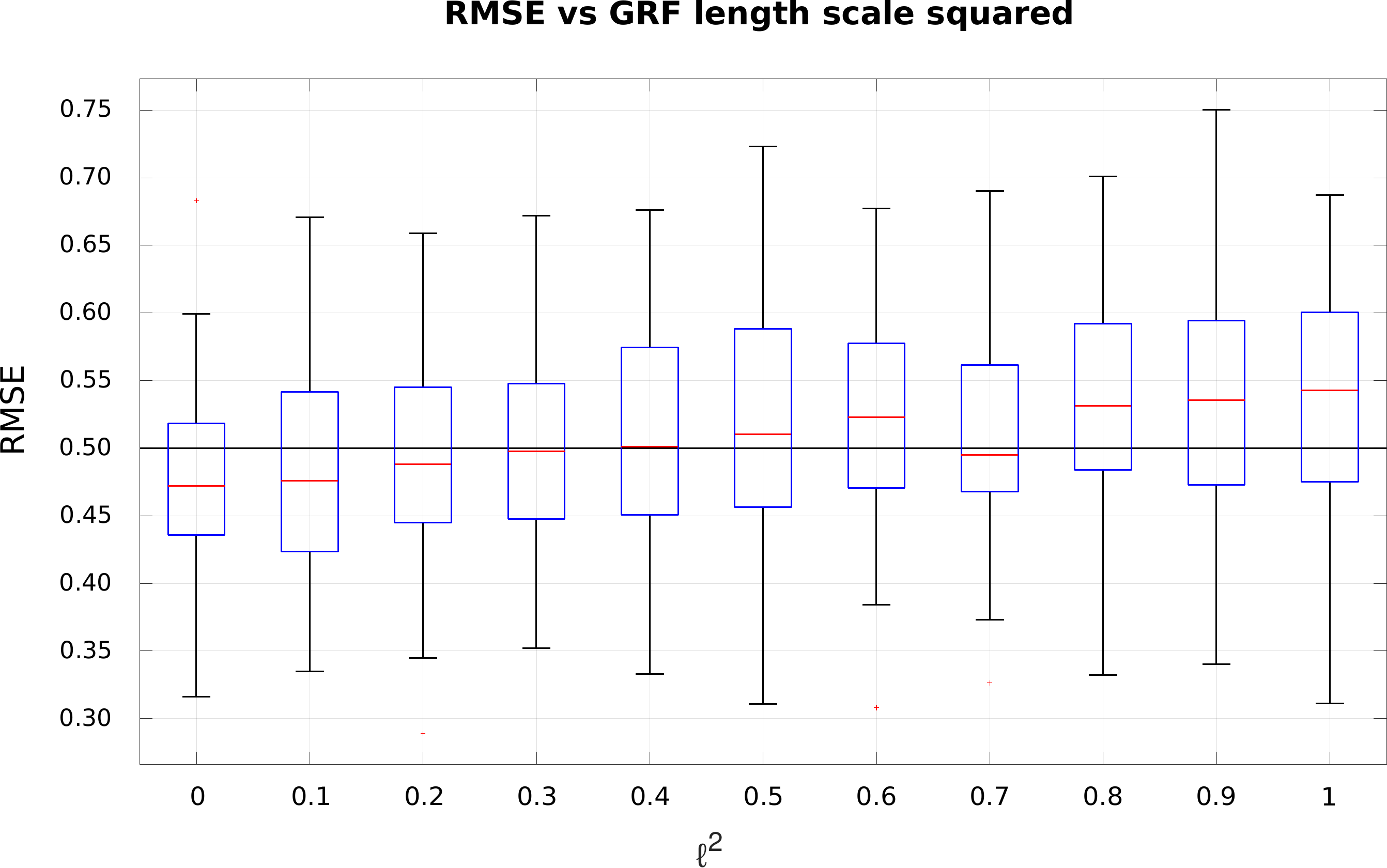}
    \caption{Root mean squared error (RMSE) between the truth and the posterior mean, using 11 different values of $\ell^2$ from $0$ to $1$.
    The first category, with $\ell^2=0$, corresponds to the uncorrelated observation error model.
    The RMSE using GRF likelihoods, i.e.~$\ell^2>0$, does not dramatically suffer in comparison to that of the white likelihood that is more common in operational practice. In exchange for this small cost in RMSE, using the GRF likelihood comes with notable gain in the accuracy of uncertainty quantification. Each box represents the middle 50\% quantile, a central line representing the median, and the whiskers span the data not considered outliers by the 1.5$\times$IQR rule. The horizontal line at 0.5 serves only to guide the eye.}\label{fig:rmse_vs_ell2}
\end{figure*}

\begin{figure*}
	\centering
    \includegraphics[width=\textwidth]{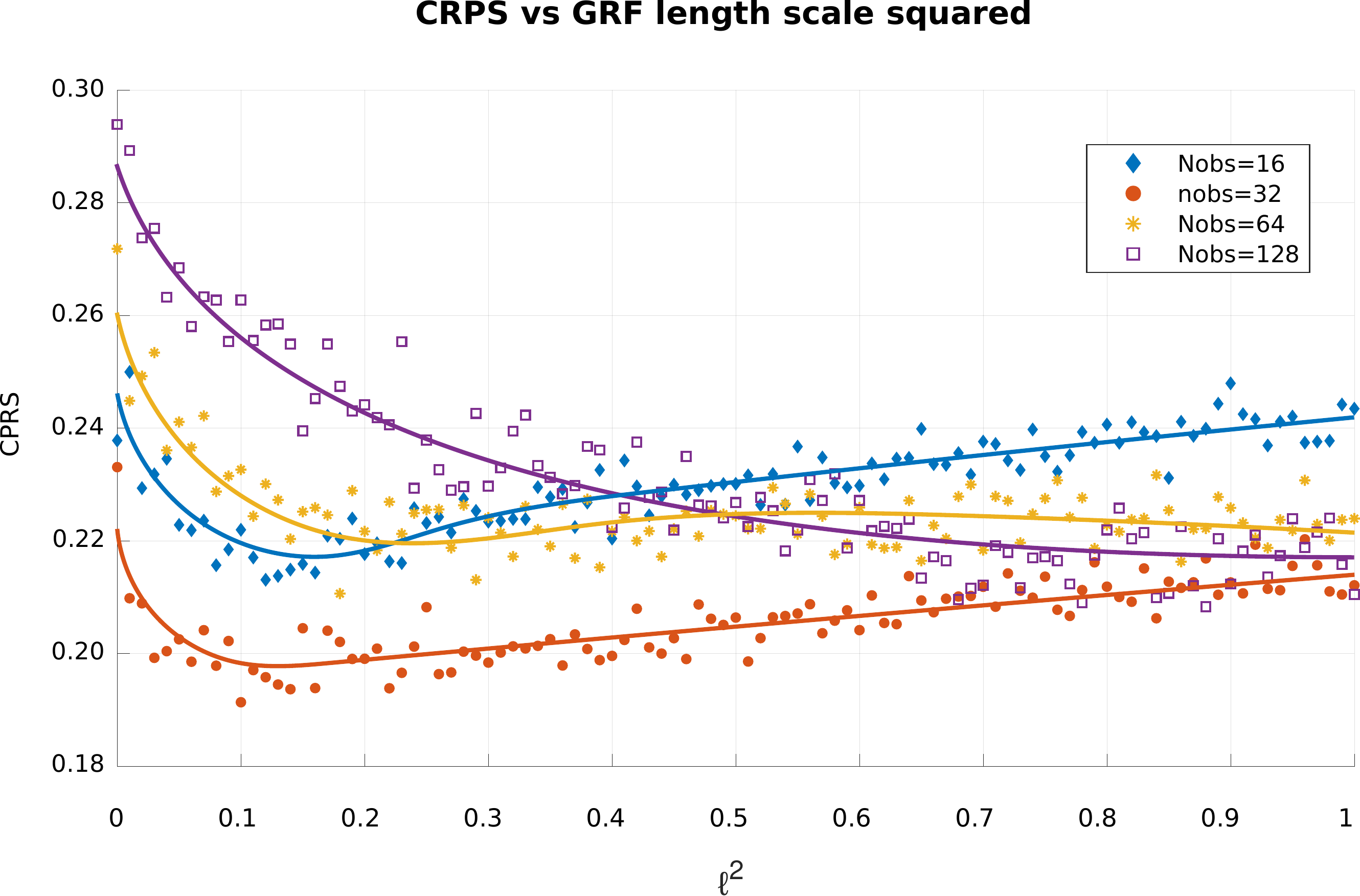}
    \caption{Continuous ranked probability score median over all time steps and grid locations, shown as a function of $\ell^2$. Each point plotted represents a particle filter assimilation run, with the same true and observed data, for different values of squared GRF length scale $\ell^2$. Each marker style represents different numbers of observations, demonstrating how the particle filter is sensitive to the number of observations. The traces are spline approximations of the data that serve to guide the eye. In each $N_{y}$ case we explored, there is a choice of $\ell^2$ that improves the particle filter CRPS. This plot emphasizes that the optimal choice of $\ell^2$ depends not only on the active scales in the underlying physics, but also on the resolution of the data. There is less information to spare about physically important scales when observations are sparse (cf. $N_{y}=16$), in which case there is only a narrow window of suitable choices for $\ell^2 \approx 0.12$ before the smoothing effect deteriorates the predictive quality of the particle filter. On the other hand, dense observations provide more abundant small-scale information that necessitates a larger choice of $\ell^2$ to achieve optimal particle filter performance. Fortunately, the more abundant information in denser observations can compensate for the injury we do to the surrogate posterior by more aggressively smoothing away small scales.} \label{fig:CRPS}
\end{figure*}

\begin{figure*}
	\centering
    \includegraphics[width=\textwidth]{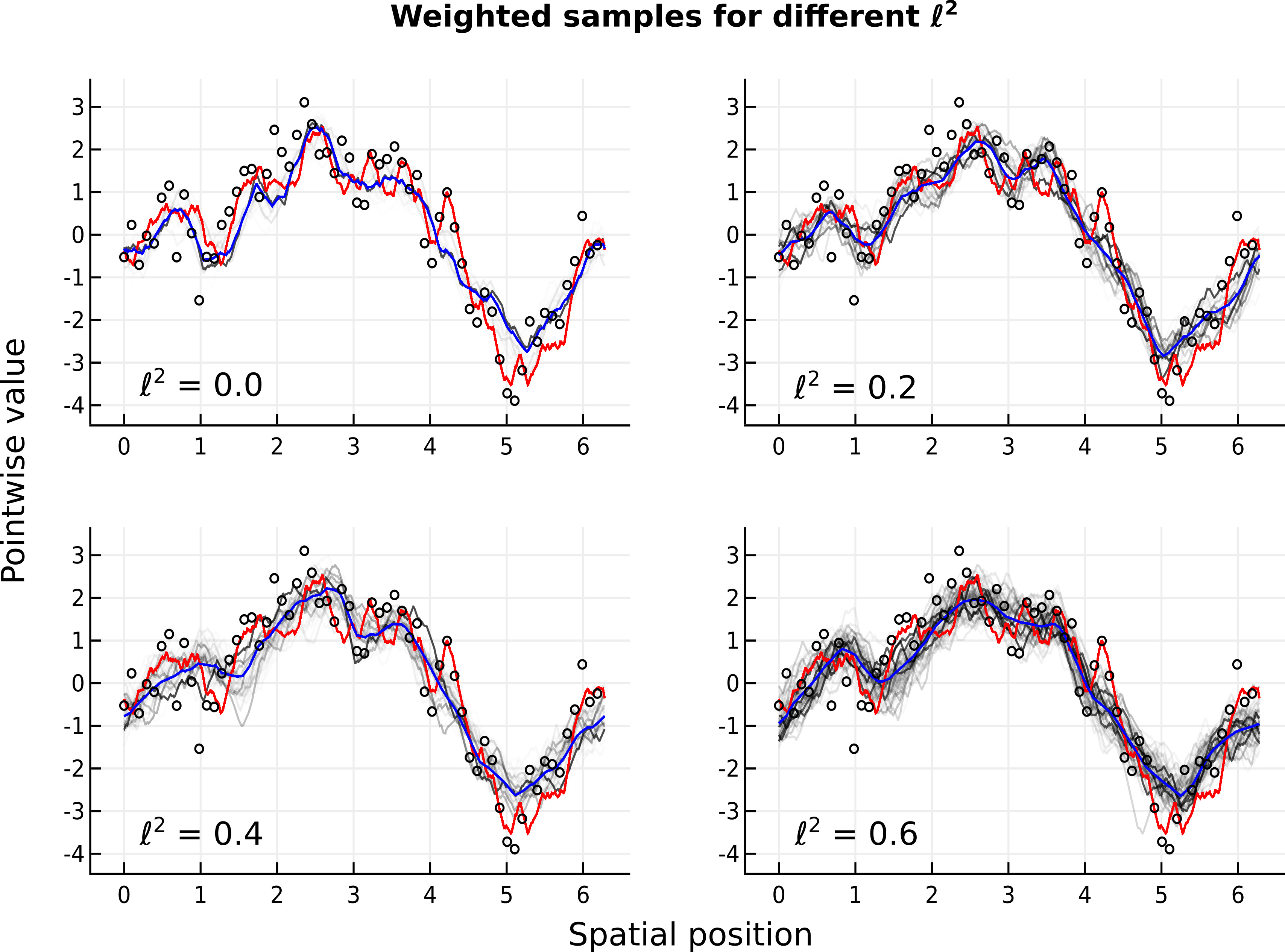}
    \caption{Pictured are the true state (red trace), PF mean (blue trace), observations (black circles), and samples from the posterior visually weighted with darkness proportional to sample weight (gray traces) for different values of $\ell^2 \in (0.0, 0.2, 0.4, 0.6)$ from left to right and top to bottom. This panel demonstrates again how a small change to the likelihood can substantially improve the problem of underestimating variance, and that this effect comes with diminishing marginal returns as the surrogate model yields progressively smoother estimates of the posterior mean.
Observe also that the samples are all realistic instantiations of the physical process, rather than overly smooth estimates.
The assimilation time shown here was chosen to exhibit monotonic improvement in $\ell^2$, which is the time-averaged behavior; due to the probabilistic nature of particle filtering, there is an abundance of times when there is not such monotonic improvement.}\label{fig:weighted_samples_by_ell2}
\end{figure*}

\begin{figure}
	\centering
    \includegraphics[width=19pc]{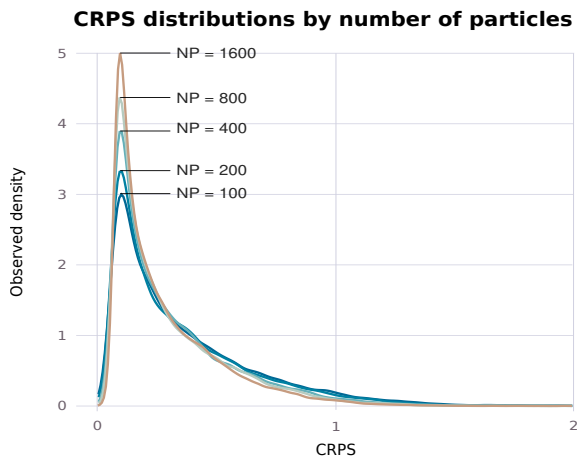}
    \caption{Kernel density estimates (KDE) of the CRPS observed for different numbers of particles demonstrate the concentration of probability as the number of particles increases while $\ell^2=0.30$ and $N_{y}=64$ are held fixed, for a fixed simulation and fixed observations thereof. Each KDE is built from the CRPS computed for each of 2048 grid cells and all 100 timesteps. The slow convergence in the number of particles is one of the reasons it is attractive to seek other means of making the particle filter more effective in sampling high-dimensional distributions.} \label{fig:crps_by_np}
\end{figure}
\end{document}